\documentclass[%
superscriptaddress,
amsmath,amssymb,
aps,
pra,
]{revtex4-2}

\usepackage{graphicx}
\usepackage{dcolumn}
\usepackage{bm}
\usepackage{hyperref}
\usepackage{braket}
\usepackage{mathtools}
\usepackage{amsthm}
\usepackage{multirow}
\usepackage{xcolor}

\DeclareMathOperator{\Arg}{Arg}
\newcommand{\inn}[2]{\langle{#1}|{#2}\rangle}

\newtheorem{Thm}{Theorem}

\newtheorem{Prop}[Thm]{Proposition}

\theoremstyle{definition}
\newtheorem{Def}[Thm]{Definition}

\begin{document}

\preprint{APS/123-QED}

\title{Three-Qubit State Preparation: Classification and Explicit Circuits}

\author{Yonghae Lee} \email{yonghaelee@kangwon.ac.kr}
\affiliation{
Department of Liberal Studies, Kangwon National University, Samcheok 25913, Republic of Korea}

\author{Taewan Kim} \email{TaewanKim@etri.re.kr} \email{Contact author: TaewanKim@etri.re.kr}
\affiliation{
Electronics and Telecommunications Research Institute, Daejeon 34129, Republic of Korea}

\date{\today}

\begin{abstract}
We present a deterministic framework for preparing an arbitrary three-qubit pure state. To leverage entanglement structure in the state-preparation task, we classify three-qubit pure states into five types with respect to a $1|2$ bipartition. Given a target state specified by its amplitudes, we provide concrete criteria and concurrence-based tests that determine its type. For each type, we derive an explicit circuit template composed of elementary single-qubit rotations and CNOT gates, with gate parameters determined systematically from the Schmidt decomposition. The full construction is described step by step from the target amplitudes, with no procedural ambiguity. As an application, we further group frequently encountered three-qubit pure states in quantum information into four classes and provide an explicit circuit for each class. Compared with prior approaches, our circuits are designed for practical use: they admit a direct algorithmic instantiation, use only CNOT gates between adjacent qubits, and for certain classes achieve smaller gate counts and circuit depth.
\end{abstract}

\maketitle


\section{Introduction}

A prescribed quantum state is a fundamental primitive across quantum information processing. In near-term noisy intermediate-scale quantum (NISQ) devices, state preparation directly competes with limited coherence times and imperfect gate fidelities, so shallow and hardware-aware preparation routines are essential~\cite{Preskill2018}. In fault-tolerant quantum computation (FTQC), state preparation remains equally central: many architectures rely on repeatedly preparing calibrated ancillary states for error correction and for promoting Clifford operations to universality~\cite{Aharonov2008,Knill1998,Bravyi2005}. These considerations motivate the design of state-preparation circuits that are not only correct, but also structured, explicit, and practically implementable.

For general $n$-qubit targets, exact state preparation is intrinsically expensive in the worst case. A generic pure state in $\mathbb{C}^{2^n}$ carries an exponential number of real degrees of freedom, and constructive synthesis methods therefore typically require $\mathcal{O}(2^n)$ elementary operations~\cite{Plesch2011}. Several decompositions realize this exponential scaling with explicit gate constructions, including schemes based on uniformly controlled rotations~\cite{Moettoenen2005}, approaches based on Shannon-style recursive decompositions~\cite{Shende2006}, and reductions derived from more general isometry decompositions~\cite{Iten2016}. Depth-optimized variants have also been investigated, emphasizing that circuit depth can be reduced while retaining asymptotically exponential gate counts~\cite{Plesch2011}. Because this exponential barrier is unavoidable in full generality, small-$n$ settings are particularly amenable to fully explicit compilation rules and to reusable circuit constructions tailored to few-qubit registers---the regime most directly relevant to NISQ experiments and to the repeated preparation of small ancilla/resource states in FTQC workflows~\cite{Plesch2011,Giraud2009,Perdomo2022}.

Among small registers, three-qubit state preparation is an important and widely studied benchmark. First, three qubits constitute the smallest setting where genuinely multipartite entanglement can occur, so preparation routines already need to accommodate entanglement features that do not arise in purely bipartite settings~\cite{Duer2000}. Second, widely used few-qubit resource states appear as subroutines in quantum communication and computation, and three-qubit ingredients such as GHZ states and graph/cluster-type states are standard examples~\cite{Bennett1993,Briegel2001,Hein2004}; this makes compact and transparent preparation circuits practically relevant~\cite{Raussendorf2001}. Third, the three-qubit case admits comparatively tractable structural descriptions; for larger numbers of qubits, entanglement classifications and related canonical descriptions quickly become more intricate, making structure-guided circuit design correspondingly less transparent~\cite{Verstraete2002}.

Several deterministic approaches to three-qubit state preparation are available in the literature. Znidari\v{c}, Giraud, and Georgeot showed that any three-qubit pure state can be prepared using at most three CNOT gates, achieving the minimum worst-case CNOT count in an all-to-all setting~\cite{Znidaric2008}. The same authors also introduced an explicit three-qubit circuit template whose parameters can be sampled to generate Haar-distributed random states, thereby providing an exact random-state generator using three CNOT gates~\cite{Giraud2009}. More recently, Perdomo \emph{et al.} proposed an alternative constructive method, including a CZ-based formulation and an explicit discussion of real-amplitude targets~\cite{Perdomo2022}.

While these works establish strong worst-case guarantees and useful universal templates, they are not always presented in a form that can be instantiated as a direct end-to-end compilation routine for a fixed hardware setting. Some intermediate choices may be left to the reader, such as selecting a particular decomposition, applying local adjustments implicitly, or relabeling the qubits. In addition, the required two-qubit-gate layout can be inconvenient under restricted connectivity. Finally, universal constructions may not exploit the additional structure of frequently encountered state families, which can lead to unnecessary parameter processing and circuit overhead in simple cases.

\begin{table*}
\caption{Summary of state-preparation circuits for three-qubit pure states. The first column indicates the entanglement types discussed in Sec.~\ref{sec:classify}. Here $R_y$ and $R_z$ denote single-qubit rotations, and $\mathrm{CNOT}$ denotes the controlled-NOT gate. We also report the total numbers of single- and two-qubit gates and the circuit depth. Since the global-phase gate $\Phi$ is physically unobservable, it is omitted from the gate counts. For the fully separable type, the reported numbers are obtained by applying the single-qubit state-preparation circuit independently to each qubit. For the biseparable type, they are obtained by combining a single-qubit preparation with a two-qubit preparation. For the SS, SE, and EE types, each circuit ends with single-qubit unitaries. For each qubit, we combine all single-qubit unitaries acting on that qubit into a single unitary and express it in the standard Z--Y--Z form in Eq.~(\ref{eq:eZYZ}); the gate counts are computed accordingly.
}
\label{tab:Summary}
\centering
\begin{tabular}{cccccccc}
\hline\hline
Type & Circuit & $n(R_y)$ & $n(R_z)$ & $n(\mathrm{CNOT})$ & Total & Depth \\
\hline
Fully separable type &Apply Fig.~\ref{fig:SP1} to each qubit &3&3&0& 6& 2 \\
Biseparable type &Combination of Figs.~\ref{fig:SP1} and~\ref{fig:SP2} &4&4&1& 9& 5 \\
SS type &Fig.~\ref{fig:SP3}(a) &5&6&2&13&6 \\
SE type &Fig.~\ref{fig:SP3}(b) &7&6&3&16&8 \\
EE type &Fig.~\ref{fig:SP3}(c) &8&7&4&19&10 \\
\hline\hline
\end{tabular}
\end{table*}

In this work we develop a fully explicit and practically oriented approach to deterministic three-qubit state preparation. Such explicit constructions also fit naturally into FTQC workflows. Many architectures repeatedly prepare small ancillary states during error correction. Our starting point is a Schmidt-decomposition viewpoint with respect to a $1|2$ bipartition of the three-qubit system, which provides a uniform way to expose the intrinsic structure of a target state and to organize the compilation problem. Among the possible choices, we use the $A|BC$ bipartition as a representative split that allows us to present our constructions in a particularly direct and uniform manner. Building on this viewpoint, we give a concrete procedure that (i) identifies the entanglement type of an arbitrary three-qubit pure state, (ii) extracts the structural data needed for synthesis, and (iii) maps this data to a state-preparation circuit through a small number of canonical circuit modules. The outcome is a structured compilation approach in which each entanglement type, and several frequently encountered subclasses, is equipped with an explicit circuit construction. A concise overview of the resulting circuit families, including gate counts and depths for each entanglement type, is given in Table~\ref{tab:Summary}.

Our constructions are designed with practical instantiation in mind. First, the overall pipeline is algorithmic and unambiguous: the circuit parameters are obtained by a clear sequence of steps from the given target amplitudes, without relying on implicit local adjustments, hidden decompositions, or qubit relabelings. Second, the resulting circuits are connectivity-aware in the sense that they are arranged to improve the CNOT layout, which can reduce the need for SWAP insertions when deploying the circuits on restricted hardware graphs typical of NISQ devices. Third, by leveraging additional structure of frequently encountered three-qubit state families, our class-specific circuits can simplify the synthesis in those regimes and, in certain cases, yield smaller circuits and shallower depth than universal constructions.

The paper is organized as follows. In Sec.~\ref{sec:classify}, we introduce a classification of three-qubit pure states based on a $1|2$ bipartition. In Sec.~\ref{sec:general}, we present a general state-preparation scheme that constructs a circuit for each entanglement type within our classification. In Sec.~\ref{sec:identify}, we provide an explicit identification procedure: given a three-qubit state specified by its amplitudes, the procedure determines its entanglement type and extracts the structural information required to instantiate the corresponding preparation circuit. In Sec.~\ref{sec:Biseparable}, we treat $A|BC$-separable targets, covering the fully separable and biseparable types; in particular, we present explicit circuits for preparing an arbitrary single-qubit state and an arbitrary two-qubit state as required building blocks. In Sec.~\ref{sec:Genuinely}, we present explicit circuits for preparing SS-, SE-, and EE-type states. In Sec.~\ref{sec:Examples}, we further group frequently encountered three-qubit pure states in quantum information into four classes and provide an explicit circuit for each class. Finally, Sec.~\ref{sec:Conclusion} summarizes our results, highlights practical advantages of our circuits, and discusses future work.

\section{Entanglement Classification with Respect to a $1|2$ Bipartition} \label{sec:classify}

In this section, we classify pure states of three qubits into five types to design state-preparation circuits for arbitrary three-qubit states. Our classification exploits a $1|2$ bipartition between one qubit and the remaining two qubits. Let $A$, $B$, and $C$ denote the three single-qubit subsystems. There are three choices, namely $A|BC$, $B|CA$, and $C|AB$. These choices are equivalent up to a relabeling of the qubits. For this reason, it is sufficient to describe the classification for the $A|BC$ bipartition.

We first separate three-qubit states into two categories using the Schmidt decomposition~\cite{Nielsen2010,Wilde2013}
across the $A|BC$ bipartition. Consider an arbitrary pure state $\ket{\psi_3}_{ABC}$ of the three-qubit system $ABC$.
Its Schmidt decomposition takes the form
\begin{equation} \label{eq:Psi3SD}
\ket{\psi_3}_{ABC}
=
\sum_{j=0}^{1}
\lambda_j \ket{\alpha_j}_A \otimes \ket{\beta_{j0}}_{BC},
\end{equation}
where the Schmidt coefficients $\lambda_j$ are nonnegative real numbers satisfying $\lambda_0^2 + \lambda_1^2 = 1$, and $\{\ket{\alpha_j}_A\}$ forms an orthonormal basis of $\mathcal{H}_A$, while $\{\ket{\beta_{j0}}_{BC}\}$ is an orthonormal set in $\mathcal{H}_{BC}$. Hereafter, we omit subsystem labels on the state vectors $\ket{\psi_3}_{ABC}$, $\ket{\alpha_j}_A$, and $\ket{\beta_{j0}}_{BC}$ whenever no confusion arises. The Schmidt rank is the number of strictly positive Schmidt coefficients. In this setting, the Schmidt rank is $1$ or $2$. If the Schmidt rank is $1$, the Schmidt decomposition reduces to a single term, i.e., a product form, and in our classification we call such states $A|BC$-separable. Without loss of generality, we order the Schmidt coefficients as $\lambda_0\ge \lambda_1$, and hence set $\lambda_1=0$ in the rank-1 case. States with Schmidt rank $2$ are called $A|BC$-entangled.

Next, we refine the above two categories into five types. The classification proceeds as follows. First, consider the $A|BC$-separable case. If $\ket{\beta_{00}}$ is separable across the $B|C$ bipartition, then the state is of the \emph{fully separable} type. Otherwise, we call it the \emph{biseparable} type.
On the other hand, the $A|BC$-entangled case is divided into three types.
To this end, we use the Schmidt decomposition in Eq.~(\ref{eq:Psi3SD}).
Specifically, we examine whether each Schmidt basis state $\ket{\beta_{j0}}$ ($j=0,1$) is separable or entangled across the $B|C$ bipartition, and group states according to the separability pattern of the pair $(\ket{\beta_{00}},\ket{\beta_{10}})$:
we speak of \emph{SS} type if both are separable, \emph{SE} type if exactly one is separable, and \emph{EE} type if both are entangled. We will refer to these three as the SS-, SE-, and EE-type $A|BC$ entanglement types.

The SS/SE/EE refinement is formulated in terms of the Schmidt basis states $\{\ket{\beta_{00}},\ket{\beta_{10}}\}$ associated with an $A|BC$ Schmidt decomposition. When the Schmidt spectrum is nondegenerate, i.e., $\lambda_0>\lambda_1$, the corresponding Schmidt basis on subsystem $A$ is unique up to phases, so the three types are well defined. In contrast, when $\lambda_0=\lambda_1$, we have $\rho_A=I_A/2$. In this case, the Schmidt basis on $A$ is not unique: any orthonormal basis of $\mathcal{H}_A$ can serve as a Schmidt basis, and the associated vectors on $BC$ can change accordingly. Throughout, in this degenerate case we \emph{define} the SS/SE/EE type by fixing the Schmidt basis on subsystem $A$ to be the computational basis $\{\ket{0}_A,\ket{1}_A\}$, and then applying the above separability test to the resulting pair $(\ket{\beta_{00}},\ket{\beta_{10}})$. We state our classification in the following definition.

\begin{Def}[type classification] \label{def:classify}
Let $\ket{\psi_3}$ be a pure state of three qubits, and consider a Schmidt decomposition across $A|BC$ as in Eq.~(\ref{eq:Psi3SD}).
In the degenerate case $\lambda_0=\lambda_1$, we fix the Schmidt basis on $A$ to be $\{\ket{0}_A,\ket{1}_A\}$.
Let $\ket{\beta_{j0}}$ denote the Schmidt vectors on system $BC$. We classify $\ket{\psi_3}$ into the following five types.
\begin{enumerate}
\item \textbf{Fully separable type:} $\ket{\psi_3}$ is $A|BC$-separable and $\ket{\beta_{00}}$ is separable across the $B|C$ bipartition.
\item \textbf{Biseparable type:} $\ket{\psi_3}$ is $A|BC$-separable and $\ket{\beta_{00}}$ is entangled across the $B|C$ bipartition.
\item \textbf{SS type:} $\ket{\psi_3}$ is $A|BC$-entangled and both $\ket{\beta_{00}}$ and $\ket{\beta_{10}}$ are separable across the $B|C$ bipartition.
\item \textbf{SE type:} $\ket{\psi_3}$ is $A|BC$-entangled and exactly one of $\ket{\beta_{00}}$ and $\ket{\beta_{10}}$ is separable across the $B|C$ bipartition.
\item \textbf{EE type:} $\ket{\psi_3}$ is $A|BC$-entangled and both $\ket{\beta_{00}}$ and $\ket{\beta_{10}}$ are entangled across the $B|C$ bipartition.
\end{enumerate}
\end{Def}

An advantage of this classification is that, to determine the type of a given three-qubit state, it suffices to compute only its Schmidt decomposition with respect to the bipartition. Since the $A|BC$ cut corresponds to a $2\times 4$ system, the Schmidt decomposition can be computed efficiently. Moreover, the entanglement type is invariant under single-qubit unitaries. Based on this property, we will provide a general scheme for three-qubit state preparation.

\section{General Preparation Scheme for Three-Qubit Pure States} \label{sec:general}

We present a general scheme for preparing arbitrary three-qubit pure states based on the Schmidt decomposition with respect to a $1|2$ bipartition, following the approach of Ref.~\cite{Murta2023}. Since our classification framework is also built on a $1|2$ bipartition, it is well suited to this approach. For consistency with Sec.~\ref{sec:classify}, we describe the scheme using the $A|BC$ bipartition. In this section, we conceptually explain how to prepare arbitrary three-qubit pure states by dividing them into two categories according to the Schmidt decomposition: the $A|BC$-separable case and the $A|BC$-entangled case. The explicit circuits for each type are derived in the next sections.

Let the target state $\ket{\psi_3}$ be assigned to one of our five types. Regardless of the type, our classification ultimately yields the Schmidt decomposition of $\ket{\psi_3}$ with respect to the $A|BC$ bipartition, as in Eq.~(\ref{eq:Psi3SD}). More explicitly, we assume that the Schmidt coefficients $\lambda_j$ and the corresponding Schmidt basis states $\ket{\alpha_j}$ and $\ket{\beta_{j0}}$ are known.

\begin{figure}
\includegraphics[clip,width=.5\columnwidth]{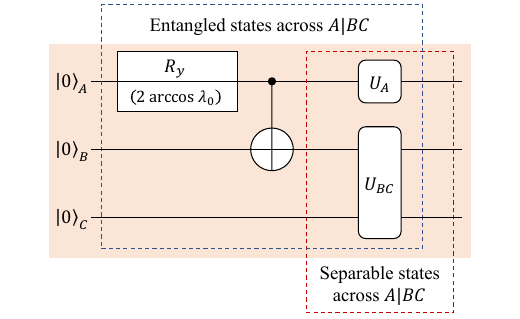}
\caption{General scheme~\cite{Murta2023} for preparing an arbitrary three-qubit pure state $\ket{\psi_3}_{ABC}$ via the Schmidt decomposition with respect to the $A|BC$ bipartition in Eq.~(\ref{eq:Psi3SD}). The unitaries $U_A$ and $U_{BC}$ are defined in Eqs.~(\ref{eq:Ua}) and~(\ref{eq:Ubc}). To instantiate this scheme, we classify three-qubit pure states into five entanglement types in Sec.~\ref{sec:classify}. If the target state is fully separable or biseparable across the $A|BC$ bipartition, only $U_A$ and $U_{BC}$ are required, whereas SS-, SE-, and EE-type $A|BC$-entangled states require the full circuit including the $R_y$ rotation and the CNOT gate.}
\label{fig:SP3General}
\end{figure}

We first consider the case where the target state is $A|BC$-entangled. Starting from the initial state $\ket{000}$, we apply the rotation $R_y(2\arccos\lambda_0)$ to qubit $A$ so that the state becomes $(\lambda_0\ket{0}+\lambda_1\ket{1}) \otimes \ket{00}$, where
\begin{equation}
R_y(\theta) = \begin{bmatrix} \cos\frac{\theta}{2} & -\sin\frac{\theta}{2} \\
\sin\frac{\theta}{2} & \cos\frac{\theta}{2} \end{bmatrix}.
\end{equation}
To create entanglement between $A$ and the subsystem $BC$, we then apply a CNOT gate with $A$ as the control and $B$ as the target, obtaining the intermediate state
\begin{equation}
\sum_{j=0}^1 \lambda_j \ket{j}_A \otimes \ket{j0}_{BC}.
\end{equation}
Finally, we apply a single-qubit unitary $U_A$ and a two-qubit unitary $U_{BC}$ defined by
\begin{eqnarray}
U_A &=& \sum_{j=0}^1 \ket{\alpha_j}\bra{j}, \label{eq:Ua} \\
U_{BC} &=& \sum_{j=0}^1 \ket{\beta_{j0}}\bra{j0} + \sum_{j=0}^1 \ket{\beta_{j1}}\bra{j1}, \label{eq:Ubc}
\end{eqnarray}
where $\{\ket{\beta_{j1}}\}_{j=0}^1$ is chosen so that $\{\ket{\beta_{jk}}\}_{j,k=0}^1$ forms an orthonormal basis of $\mathcal{H}_{BC}$; then $U_{BC}$ is unitary by construction. This procedure prepares the target state exactly. Figure~\ref{fig:SP3General} illustrates the corresponding conceptual circuit. The concrete implementation of $U_{BC}$ for the SS-, SE-, and EE-types will be discussed later.

Once the preparation procedure for $A|BC$-entangled states is understood, the circuit for $A|BC$-separable states follows immediately. In the $A|BC$-separable case, the target state is a product across the cut; hence neither the CNOT gate nor the $R_y$ rotation is required, and one directly applies $U_A$ and $U_{BC}$ to $\ket{000}$, as shown in Fig.~\ref{fig:SP3General}. In other words, state preparation for fully separable and biseparable states reduces to a combination of single-qubit and two-qubit state-preparation procedures.

\section{Identification of Three-Qubit Entanglement Types from State Coefficients} \label{sec:identify}

In this section, we provide a systematic decision procedure to identify the entanglement type of a three-qubit pure state directly from its amplitudes. Since our type classification is defined through Schmidt decompositions with respect to appropriate bipartitions, the main task is to extract the corresponding Schmidt ranks and Schmidt coefficients from the amplitudes.

Any target three-qubit pure state can be specified by its amplitudes in the computational basis as
\begin{equation} \label{eq:Psi3}
\ket{\psi_3}
\coloneqq
c_0\ket{000} +c_1\ket{001} +c_2\ket{010} +c_3\ket{011} +c_4\ket{100} +c_5\ket{101} +c_6\ket{110} +c_7\ket{111},
\end{equation}
where the amplitudes $c_j$ satisfy the normalization condition, i.e., $\sum_{j=0}^7|c_j|^2=1$; here $\{\ket{0},\ket{1}\}$ denotes the computational basis of each single-qubit Hilbert space. Now we describe how to determine the type of $\ket{\psi_3}$ according to our classification scheme. As a first step, for the $A|BC$ bipartition, we must decide whether the given state is $A|BC$-separable or $A|BC$-entangled. Instead of computing the Schmidt decomposition (or, more directly, the spectral decomposition of the reduced state), we test $A|BC$ separability by checking whether the target state can be written in a product form between the single-qubit system $A$ and the two-qubit system $BC$. This approach yields an $A|BC$-separability test with a relatively small computational overhead. More specifically, we present the separability conditions in the following proposition.

\begin{Prop}[$A|BC$ separability] \label{prop:conditions}
Let $\ket{\psi_3}$ be the three-qubit pure state in Eq.~\eqref{eq:Psi3}. Then it is $A|BC$-separable if and only if one of the following three conditions holds:
\begin{enumerate}
\item $c_0=c_1=c_2=c_3=0$.
\item $c_4=c_5=c_6=c_7=0$.
\item The coefficient vectors $(c_0,c_1,c_2,c_3)$ and $(c_4,c_5,c_6,c_7)$ are proportional; equivalently, there exists a nonzero $k\in\mathbb{C}$ such that $c_{j+4}=kc_j \quad (j=0,1,2,3)$.
\end{enumerate}
\end{Prop}

Note that, in Proposition~\ref{prop:conditions}, the first and second conditions correspond to degenerate cases where subsystem \(A\) is fixed to \(\ket{1}\) or \(\ket{0}\), respectively. If the third condition holds, then for any $j'\in\{0,1,2,3\}$ with $c_{j'}\neq 0$, $\ket{\psi_3}$ admits the following product form:
\begin{equation} \label{eq:Biseparable3}
\ket{\psi_3}=
\left( \frac{1}{\tau} \left( c_{j'} \ket{0} + c_{{j'}+4} \ket{1} \right) \right)
\otimes
\left( \frac{\tau}{c_{j'}} \left( c_0 \ket{00} + c_1 \ket{01} + c_2 \ket{10} + c_3 \ket{11} \right) \right),
\end{equation}
where $\tau \coloneqq \sqrt{|c_{j'}|^2+|c_{{j'}+4}|^2}$ is real and positive. In particular, $\ket{\psi_3}$ is $A|BC$-entangled if and only if none of the above conditions holds.

\subsection{$A|BC$-separable case} \label{sec:ABCsep}

If one of the three conditions in Proposition~\ref{prop:conditions} holds, then the target state $\ket{\psi_3}$ is $A|BC$-separable. To determine whether the state is fully separable, we further decide whether the two-qubit subsystem $BC$ is entangled. For this purpose, we employ the concurrence~\cite{Hill1997,Wootters1998}. Let the two-qubit state on $BC$ be denoted by $\ket{\psi_2}_{BC}$, which can be written in the computational basis as
\begin{equation} \label{eq:Psi2} 
\ket{\psi_2}_{BC} \coloneqq d_0 \ket{00}_{BC} +d_1 \ket{01}_{BC} + d_2 \ket{10}_{BC}+ d_3 \ket{11}_{BC},
\end{equation}
where $\sum_{j=0}^3 |d_j|^2 = 1$. The \emph{concurrence} $\mathcal{C}$ of a two-qubit pure state $\ket{\psi_2}$ is defined as
\begin{equation} \label{eq:Concurrence}
\mathcal{C} \coloneqq 2 \left| d_0 d_3 - d_1 d_2 \right|.
\end{equation}
If $\mathcal{C}=0$, then $\ket{\psi_2}$ is separable; otherwise, if $\mathcal{C}>0$, it is entangled. 
Therefore, by evaluating the concurrence of $\ket{\psi_2}$, we can decide whether the target state $\ket{\psi_3}$ is fully separable or biseparable.

\begin{enumerate}
\item \textbf{Fully separable type.}
If \(\mathcal{C}=0\), then \(\ket{\psi_2}\) is $B|C$-separable and hence the target state \(\ket{\psi_3}\) is fully separable. In this case, if \(d_0\neq 0\), one may take
\begin{equation} \label{eq:BCProduct}
\ket{\psi_2} =
\left( \frac{1}{\sqrt{|d_0|^2+|d_2|^2}} \left( d_0 \ket{0} + d_2 \ket{1} \right) \right)
\otimes
\left( \frac{1}{\sqrt{|d_0|^2+|d_1|^2}} \left( d_0 \ket{0} + d_1 \ket{1} \right) \right),
\end{equation}
where the equality follows from the condition \(d_0 d_3=d_1 d_2\). Likewise, if \(d_0=0\) then \(\mathcal{C}=0\) implies \(d_1 d_2=0\), and \(\ket{\psi_2}\) reduces to an obvious product state supported on a computational-basis subspace. Consequently, in the fully separable case, the preparation of \(\ket{\psi_3}\) reduces to applying single-qubit state preparation independently on each qubit system, as described in Sec.~\ref{sec:Biseparable}.

\item \textbf{Biseparable type.}
If \(\mathcal{C}>0\), then \(\ket{\psi_2}\) is \(B|C\)-entangled and hence the target state \(\ket{\psi_3}\) is biseparable. In this case, the target state can be written as the product of a single-qubit state on system \(A\) and a two-qubit state \(\ket{\psi_2}\) on system \(BC\). Therefore, the preparation of a biseparable three-qubit state can be carried out by combining single-qubit state preparation with two-qubit state preparation. In Sec.~\ref{sec:Biseparable}, we provide an explicit circuit for preparing an entangled two-qubit pure state; the construction is based on the Schmidt decomposition of the two-qubit state.
\end{enumerate}

Since the goal of the present section is to provide a systematic decision procedure, it is useful to describe how to obtain the Schmidt decomposition of the two-qubit state \(\ket{\psi_2}\) in Eq.~\eqref{eq:Psi2}. Ref.~\cite{Lee2025} provides analytical formulas for the Schmidt decomposition of an arbitrary two-qubit pure state. As shown therein, there is no single closed-form expression that applies to all two-qubit pure states. Instead, the appropriate formula depends on the structure of the state's coefficients. To classify the cases, we define the \textit{diagonal} coefficient $\mathcal{D}$~\cite{Lee2025} of $\ket{\psi_2}$ as
\begin{equation}
\mathcal{D} \coloneqq d_0^* d_1 + d_2^* d_3,
\label{eq:Diagonal}
\end{equation}
where \((\cdot)^*\) denotes complex conjugation. Depending on whether \(\mathcal{D}\) is zero or nonzero, different formulas apply.

We first treat the generic case $\mathcal{D}\neq 0$, which holds for almost all entangled two-qubit states. To apply the Schmidt decomposition formulas, we compute the quantum concurrence $\mathcal{C}$ in Eq.~(\ref{eq:Concurrence}) and the diagonal coefficient $\mathcal{D}$ in Eq.~(\ref{eq:Diagonal}). Using these, for each $j=0,1$, we define the following coefficients:
\begin{align}
\tau_j &\coloneqq \left(\frac{1 + (-1)^j \sqrt{1 - \mathcal{C}^2}}{2} \right)^{1/2}, \\
B_j &\coloneqq \tau_j^2 - |d_0|^2 - |d_3|^2, \\
E_j &\coloneqq d_0 \mathcal{D} + d_1 B_j, \\
F_j &\coloneqq d_2 \mathcal{D} + d_3 B_j.
\end{align}
Following the convention of Ref.~\cite{Lee2025}, the Schmidt decomposition of \(\ket{\psi_2}\) is then given by
\begin{equation} \label{eq:SDNonDiagonal}
\ket{\psi_2}
= \sum_{j=0}^1 \tau_j \left( \frac{1}{\sqrt{|E_j|^2+|F_j|^2}} \left( E_j \ket{0} + F_j \ket{1} \right) \right)
\otimes \left( \frac{1}{\sqrt{|\mathcal{D}|^2+|B_j|^2}} \left( \mathcal{D}^* \ket{0} + B_j \ket{1} \right) \right).
\end{equation}
In contrast, when the diagonal coefficient of the target state is zero, i.e., $\mathcal{D} = 0$, the state admits a simpler form of Schmidt decomposition~\cite{Lee2025}:
\begin{equation} \label{eq:SDDiagonal}
\ket{\psi_2}_{BC} = \sum_{j=0}^1 \tau_j \left( \frac{1}{\tau_j} \left( d_j \ket{0} + d_{j+2} \ket{1} \right) \right) \otimes \ket{j}.
\end{equation}
In this case, the Schmidt coefficients reduce to $\tau_j = \sqrt{|d_j|^2 + |d_{j+2}|^2}$. This dichotomy can be understood from the reduced state on a single qubit. To obtain the Schmidt decomposition across \(B|C\), one may compute the spectral decomposition of the \(2\times 2\) reduced density matrix of \(B\) (or \(C\)). When \(\mathcal{D}=0\), the off-diagonal entries of this reduced density matrix vanish, so it becomes diagonal in the computational basis and its eigen-decomposition is immediate. Consequently, the Schmidt decomposition takes the simple form in Eq.~\eqref{eq:SDDiagonal}.

\subsection{$A|BC$-entangled case} \label{sec:ABCent}

We turn to the case where the target state is \(A|BC\)-entangled, and explain how to further refine its type within our classification. In order to follow the general preparation scheme introduced in Sec.~\ref{sec:general}, we compute the Schmidt decomposition with respect to the \(A|BC\) bipartition. To this end, for the target state \(\ket{\psi_3}\) in Eq.~\eqref{eq:Psi3}, we first evaluate the reduced state on subsystem \(A\), denoted by \(\rho_A\), as
\begin{equation}\label{eq:rhoA}
\rho_A =
 \left( \sum_{j=0}^{3} |c_j|^2 \right) \ket{0}\bra{0}
+ \left( \sum_{j=0}^{3} c_j c_{j+4}^* \right) \ket{0}\bra{1}
+ \left( \sum_{j=0}^{3} c_j^* c_{j+4} \right) \ket{1}\bra{0}
+ \left( \sum_{j=4}^{7} |c_j|^2 \right) \ket{1}\bra{1}.
\end{equation}

Next, we compute the spectral decomposition of \(\rho_A\). Since an analytic formula for the eigendecomposition of a \(2\times 2\) Hermitian matrix is standard, we omit the details and assume that the following decomposition is known:
\begin{equation}
\rho_A = \sum_{j=0}^1 \lambda_j^2 \ket{\alpha_j}\bra{\alpha_j},
\end{equation}
where the eigenvalues \(\lambda_j^2\) satisfy \(\lambda_0^2+\lambda_1^2=1\), and the eigenvectors \(\{\ket{\alpha_0},\ket{\alpha_1}\}\) form an orthonormal basis of the single-qubit Hilbert space of subsystem \(A\). To align the notation with the Schmidt decomposition introduced earlier, we assume \(\lambda_0\ge \lambda_1\). This provides the Schmidt coefficients \(\lambda_j\) and the Schmidt basis states \(\ket{\alpha_j}\) on subsystem \(A\).

To obtain the Schmidt basis states on subsystem \(BC\), we form the rank-one projector onto \(\ket{\alpha_0}\) and apply it to the target state. This yields the first term of the Schmidt decomposition. Since the resulting vector is, by construction, \(A|BC\)-separable, we can use the known \(\lambda_0\) and \(\ket{\alpha_0}\) to determine the corresponding first Schmidt basis state on subsystem \(BC\), denoted by \(\ket{\beta_{00}}\). Altogether, this step can be summarized by the identity
\begin{equation}
\left(\ket{\alpha_0}\bra{\alpha_0}\right)_A \otimes I_{BC}\ket{\psi_3}_{ABC}
=
\lambda_0\ket{\alpha_0}_A \otimes \ket{\beta_{00}}_{BC},
\end{equation}
where \(I_{BC}\) denotes the identity operator on subsystem \(BC\).
To obtain the second Schmidt basis state on subsystem \(BC\), we subtract the first Schmidt term from \(\ket{\psi_3}\). Since \(\lambda_1\) and \(\ket{\alpha_1}\) are also known, the second Schmidt basis state \(\ket{\beta_{10}}\) can be determined in the same manner. Equivalently, we may write
\begin{equation}
\ket{\psi_3}_{ABC} - \lambda_0\ket{\alpha_0}_A \otimes \ket{\beta_{00}}_{BC}
=
\lambda_1\ket{\alpha_1}_A \otimes \ket{\beta_{10}}_{BC}.
\end{equation}

In principle, one might try to obtain the $BC$-side Schmidt basis by diagonalizing the reduced state $\rho_{BC}$. However, this approach is insufficient for our purposes because the Schmidt decomposition requires a phase-consistent pairing between the two local bases $\ket{\alpha_j}$ and $\ket{\beta_{j0}}$. While the eigenvectors of $\rho_A$ determine $\ket{\alpha_j}$ only up to independent global phase factors, those phase choices fix the corresponding phases of $\ket{\beta_{j0}}$ in the decomposition. In contrast, extracting $\ket{\beta_{j0}}$ from $\rho_{BC}$ loses this information: eigenvectors of $\rho_{BC}$ are likewise defined only up to arbitrary phases, and the reduced state contains no record of how the phases should be matched to the previously chosen $\ket{\alpha_j}$. Consequently, once $\ket{\alpha_j}$ is fixed from $\rho_A$, we determine $\ket{\beta_{j0}}$
 via projection onto $\ket{\alpha_j}$, which preserves the correct phase relationship and yields Schmidt basis states on $BC$ that are consistent with the chosen $\ket{\alpha_j}$.

Once the Schmidt basis states \(\{\ket{\beta_{00}},\ket{\beta_{10}}\}\) have been determined, we further classify the target state into one of the SS, SE, and EE types by evaluating the concurrence~\cite{Hill1997,Wootters1998} of these two-qubit states.

\begin{enumerate}
\item \textbf{SS type.}
If both vectors \(\ket{\beta_{00}}\) and \(\ket{\beta_{10}}\) have zero concurrence, then the target state is of SS type. In this case, we compute the corresponding product forms of the two vectors using Eq.~\eqref{eq:BCProduct}. The design of an explicit preparation circuit for an SS-type target state is described in Sec.~\ref{sec:SS}.

\item \textbf{SE type.}
If exactly one of \(\ket{\beta_{00}}\) and \(\ket{\beta_{10}}\) has zero concurrence while the other has nonzero concurrence, then the target state is of SE type. In this case, we compute the product form of the \(B|C\)-separable vector using Eq.~\eqref{eq:BCProduct}. We explain how to design an explicit preparation circuit for the SE type in Sec.~\ref{sec:SE}.

\item \textbf{EE type.}
If both vectors have nonzero concurrence, then the target state is of EE type. In this case, we compute the Schmidt decomposition of the first vector \(\ket{\beta_{00}}\) using the analytic formulas for two-qubit pure states presented in Sec.~\ref{sec:ABCsep}. The corresponding circuit design is provided in Sec.~\ref{sec:EE}.
\end{enumerate}

In this section, we presented a systematic procedure for identifying the entanglement type of an arbitrary three-qubit pure state. Moreover, for each type, we described explicit formulas and methods to extract the relevant structural data, including the Schmidt coefficients and Schmidt basis states. In the subsequent sections, we use this information to construct state-preparation circuits tailored to each entanglement type.

\section{Preparation of $A|BC$-Separable Three-Qubit Pure States: Reduction to Single- and Two-Qubit State Preparation}
\label{sec:Biseparable}

In this section, we present a constructive method for designing circuits that prepare biseparable three-qubit pure states, encompassing both the fully separable and the biseparable classes. Focusing on states that are separable with respect to the $A|BC$ bipartition, any target state can be written such that the subsystem $A$ is a single-qubit pure state, while the subsystem $BC$ is either a product of two single-qubit pure states or an entangled two-qubit pure state. Accordingly, we first describe an explicit circuit for preparing an arbitrary single-qubit pure state, and then present circuits for preparing arbitrary entangled two-qubit pure states on $BC$.

\subsection{Exact single-qubit state preparation} \label{sec:SingleQubit}

We first present an explicit single-qubit circuit for the mathematically exact preparation of an arbitrary pure state from the computational-basis state $\ket{0}$, with gate parameters determined directly by the state coefficients. We consider an arbitrary single-qubit state $\ket{\psi_1}$, given by
\begin{equation} \label{eq:Psi1}
\ket{\psi_1} \coloneqq a \ket{0} + b \ket{1},
\end{equation}
where $|a|^2+|b|^2=1$. Conceptually, the task of single-qubit state preparation is to set the magnitudes $|a|$ and $|b|$ and the relative phase $\Arg(b)-\Arg(a)$. An $R_y$ rotation is applied to create a superposition of $\ket{0}$ and $\ket{1}$ and to fix the magnitudes of the two amplitudes. An $R_z$ rotation then incorporates the desired relative phase. Thus, the target state can be obtained via
\begin{equation}
\ket{\psi_1}
= R_z \left(\Arg(b)-\Arg(a) \right) R_y \left(2\arccos|a| \right) \Phi \left(\Arg(b)+\Arg(a) \right) \ket{0},
\end{equation}
where, as usual, the rightmost gate acts first. The global phase gate $\Phi$ is included only to indicate mathematically exact state preparation and can be omitted in practice, since a global phase is physically unobservable. Here the single-qubit rotation $R_z$ and the global phase gate $\Phi$ are given by
\begin{equation}
R_z(\theta) = \begin{bmatrix} e^{-i\theta/2} & 0 \\ 0 & e^{i\theta/2} \end{bmatrix} \quad \mathrm{and} \quad
\Phi(\theta) = \begin{bmatrix} e^{i\theta/2} & 0 \\ 0 & e^{i\theta/2} \end{bmatrix}.
\end{equation}

\begin{figure}
\includegraphics[clip,width=.5\columnwidth]{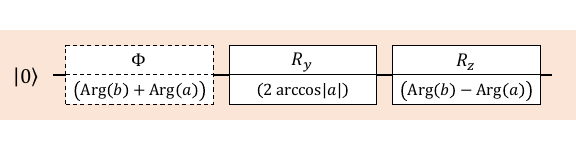}
\caption{Circuit for preparing an arbitrary single-qubit state $\ket{\psi_1}$ as in Eq.~(\ref{eq:Psi1}). This circuit serves as a building block for preparing biseparable three-qubit pure states. Since a global phase has no observable physical effect, $\Phi$ is introduced only to indicate mathematically exact state preparation; it is therefore drawn with a dashed outline and can be omitted in practice. The same building blocks also suffice to prepare fully separable three-qubit pure states.}
\label{fig:SP1}
\end{figure}

Figure~\ref{fig:SP1} shows the circuit for preparing arbitrary single-qubit states, parametrized by their coefficients. While an arbitrary single-qubit unitary in $\mathrm{SU}(2)$ generally requires three rotations in a Z--Y--Z decomposition, the present task requires fewer gates: as illustrated in Fig.~\ref{fig:SP1}, two single-qubit gates suffice up to a global phase. This simplification is possible because an arbitrary unitary must specify the images of both computational-basis states, whereas state preparation only needs to fix the image of $\ket{0}$.

\subsection{Exact two-qubit state preparation} \label{sec:TwoQubit}

We next construct a circuit for preparing an arbitrary two-qubit pure state $\ket{\psi_2}$ given by Eq.~(\ref{eq:Psi2}), assuming $C>0$ to exclude the separable case. The two-qubit state then admits the Schmidt decomposition:
\begin{equation} \label{eq:Psi2SD}
\ket{\psi_2} = \sum_{j=0}^1 \tau_j \ket{\gamma_j} \otimes\ket{\delta_j},
\end{equation}
where the Schmidt coefficients $\tau_j$ are nonnegative real numbers satisfying $\tau_0^2 + \tau_1^2 = 1$, and $\{\ket{\gamma_j}\}$ and $\{\ket{\delta_j}\}$ are orthonormal bases for the single-qubit subsystems $B$ and $C$, respectively.

In line with the general strategy introduced in Sec.~\ref{sec:classify}, we use this Schmidt decomposition as the starting point for constructing a preparation circuit. The state preparation starts from the computational-basis state $\ket{00}$, and the first step is to apply the rotation $R_y(2\arccos\tau_0)$ to qubit $B$ in order to encode the Schmidt coefficients. A CNOT gate is then applied, with qubit $B$ as control and qubit $C$ as target, resulting in the intermediate entangled state
\begin{equation}
\tau_0 \ket{00}_{BC} + \tau_1 \ket{11}_{BC}.
\end{equation}
Let us consider the single-qubit unitaries $V_B$ and $V_C$ defined as
\begin{equation}
V_B \coloneqq \sum_{j=0}^1 \ket{\gamma_j}\bra{j} \quad \mathrm{and} \quad
V_C \coloneqq \sum_{j=0}^1 \ket{\delta_j}\bra{j}. \label{eq:VbVc}
\end{equation}
By applying the single-qubit unitaries $V_B$ and $V_C$ to qubits $B$ and $C$, respectively, the two-qubit state $\ket{\psi_2}$ is obtained.

To determine the unitaries $V_B$ and $V_C$ in Eq.~(\ref{eq:VbVc}) corresponding to this Schmidt decomposition, we consider an arbitrary single-qubit unitary matrix
\begin{equation}
U= a \ket{0}\bra{0} + b \ket{0}\bra{1} + c \ket{1}\bra{0} + d \ket{1}\bra{1}.
\end{equation}
Based on the decomposition proposed in Ref.~\cite{Barenco1995}, the unitary $U$ with $a\neq 0$ is implemented as follows:
\begin{equation} \label{eq:eZYZ}
U = 
\Phi\left(\Arg(a)+\Arg(d)\right)
R_z\left(\Arg(a^*)+\Arg(c)\right)
R_y\left(2\arccos|a|\right)
R_z\left(\Arg(c^*)+\Arg(d)\right),
\end{equation}
where the decomposition is parametrized by $|a|$ and the arguments of $a$, $b$, $c$, and $d$, with the overall phase handled via $\Phi$. For the exceptional case $a=0$, it can instead be implemented as
\begin{equation}
U =
\Phi\left(\Arg(b)+\Arg(c)-\pi\right)
R_z\left(\Arg(b^*)+\Arg(c) + \pi\right)
R_y\left(\pi\right).
\end{equation}
Since the set of unitaries with $a=0$ forms a measure-zero subset of $U(2)$, in this work we regard Eq.~(\ref{eq:eZYZ}) as describing the generic case and focus on the regime $a\neq 0$. The unitaries $V_B$ and $V_C$ in Eq.~(\ref{eq:VbVc}) are obtained by instantiating $U$ in Eq.~(\ref{eq:eZYZ}) with the corresponding parameters.

\begin{figure*}
\centering
\includegraphics[width=\textwidth]{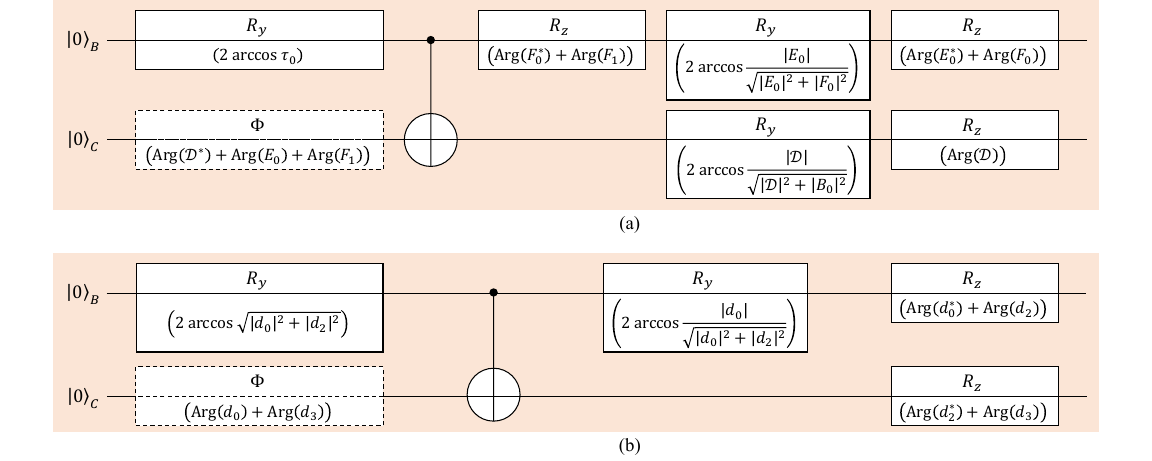}
\caption{Circuit for preparing an arbitrary two-qubit state $\ket{\psi_2}$ as in Eq.~(\ref{eq:Psi2}), equivalently expressed in the Schmidt form of Eq.~(\ref{eq:Psi2SD}). This circuit serves as a building block for preparing biseparable three-qubit pure states. We assume that the two-qubit state is entangled with respect to the $B|C$ bipartition.
The circuit constructions are divided into two cases according to the diagonal coefficient $\mathcal{D}$ in Eq.~(\ref{eq:Diagonal}):
(a) generic case $\mathcal{D}\neq 0$, where, to simplify the gate parametrization, the single-qubit unitaries $V_B$ and $V_C$ are expressed in terms of the Schmidt coefficients $\tau_j$ and the auxiliary quantities $B_j$, $E_j$, and $F_j$ defined in Eq.~(\ref{eq:SDNonDiagonal});
(b) special case $\mathcal{D}=0$, where the simpler Schmidt decomposition in Eq.~(\ref{eq:SDDiagonal}) allows the gate parameters to be represented directly in terms of the amplitudes $d_j$ in Eq.~(\ref{eq:Psi2}).}
\label{fig:SP2}
\end{figure*}

By combining the steps above, we obtain explicit preparation circuits for arbitrary entangled two-qubit pure states. Since the analytic formulas used to derive the Schmidt decomposition depend on the diagonal coefficient $\mathcal{D}$ defined in Eq.~(\ref{eq:Diagonal}), we distinguish two cases. If an entangled two-qubit pure state satisfies $\mathcal{D}\neq 0$, we use the Schmidt decomposition in Eq.~(\ref{eq:SDNonDiagonal}), which determines the single-qubit unitaries $V_B$ and $V_C$ in Eq.~(\ref{eq:VbVc}). The corresponding state-preparation circuit is shown in Fig.~\ref{fig:SP2}(a). On the other hand, if $\mathcal{D}=0$, we instead use the simpler Schmidt decomposition in Eq.~(\ref{eq:SDDiagonal}), which yields a simpler form of $V_B$ and $V_C$. In particular, the single-qubit unitary $V_B$ takes the form
\begin{equation}
V_B = \frac{d_0}{\tau_0} \ket{0}\bra{0} + \frac{d_1}{\tau_1} \ket{0}\bra{1} + \frac{d_2}{\tau_0} \ket{1}\bra{0} + \frac{d_3}{\tau_1} \ket{1}\bra{1},
\end{equation}
and $V_C$ reduces to the identity operator on subsystem $C$. The complete circuit for this case is shown in Fig.~\ref{fig:SP2}(b).

Note that we have presented explicit circuit constructions for preparing arbitrary entangled two-qubit pure states with specified amplitudes, by deriving coefficient-dependent Schmidt decompositions and tailoring the single-qubit unitaries $V_B$ and $V_C$. Whereas implementing an arbitrary two-qubit unitary generally requires three CNOT gates~\cite{Vidal2004,Vatan2004}, preparing an entangled two-qubit pure state in our scheme requires only a single CNOT gate, and is therefore cheaper in terms of entangling resources.

\section{Preparation of $A|BC$-Entangled Three-Qubit Pure States} \label{sec:Genuinely}

In this section, we present explicit state-preparation circuits for $A|BC$-entangled target states. Specifically, we show how to realize the general scheme of Sec.~\ref{sec:general} for the SS-, SE-, and EE-type families. Following the discussion in Sec.~\ref{sec:ABCent}, we assume that the Schmidt data of the target state are given. For each type, we (i) derive a convenient canonical form of the Schmidt basis states, (ii) determine the essential action of the two-qubit unitary $U_{BC}$ in Fig.~\ref{fig:SP3General}, and (iii) construct an explicit circuit that prepares an arbitrary $A|BC$-entangled three-qubit pure state of that type.

Technically, the main task is to implement the required two-qubit transformation $U_{BC}$ using elementary single-qubit gates and CNOT gates. While one could in principle invoke standard decompositions for arbitrary two-qubit unitaries~\cite{Vidal2004,Vatan2004,Moettoenen2004,Iten2016}, they are not well suited to our purpose. The constructions in Refs.~\cite{Vidal2004,Vatan2004}, for instance, are tailored to implement a given unitary on the entire computational basis and rely on specific two-qubit decompositions~\cite{Khaneja2001,Kraus2001,Zhang2003}, which makes them unnecessarily heavy in our setting. For state preparation, however, it suffices to control the action of $U_{BC}$ on two orthogonal input states---the Schmidt basis states $\ket{\beta_{00}}$ and $\ket{\beta_{10}}$. We therefore build the needed two-qubit mappings directly, by combining standard circuit techniques from quantum computation~\cite{Nielsen2010,Barenco1995}, which yields a unified and transparent treatment of the SS-, SE-, and EE-type families.

\subsection{State-preparation circuit for SS-type states} \label{sec:SS}

We present an explicit circuit implementation for preparing an SS-type target state $\ket{\psi_3}$. Recall that a state is of SS type if, in the Schmidt decomposition with respect to the $A|BC$ bipartition in Eq.~(\ref{eq:Psi3SD}), the Schmidt basis vectors $\ket{\beta_{00}}$ and $\ket{\beta_{10}}$ are separable with respect to the $B|C$ bipartition. Specifically, for each $j=0,1$, suppose that
\begin{equation} \label{eq:SS}
\ket{\beta_{j0}} = \ket{\gamma_j} \otimes \ket{\delta_j},
\end{equation}
where $\ket{\gamma_j}$ and $\ket{\delta_j}$ are arbitrary single-qubit pure states on systems $B$ and $C$, respectively. Since $\ket{\beta_{00}}$ and $\ket{\beta_{10}}$ are orthogonal, at least one of the pairs $\{\ket{\gamma_j}\}_{j=0,1}$ and $\{\ket{\delta_j}\}_{j=0,1}$ must consist of orthogonal single-qubit states.

\begin{figure*}
\centering
\includegraphics[width=\textwidth]{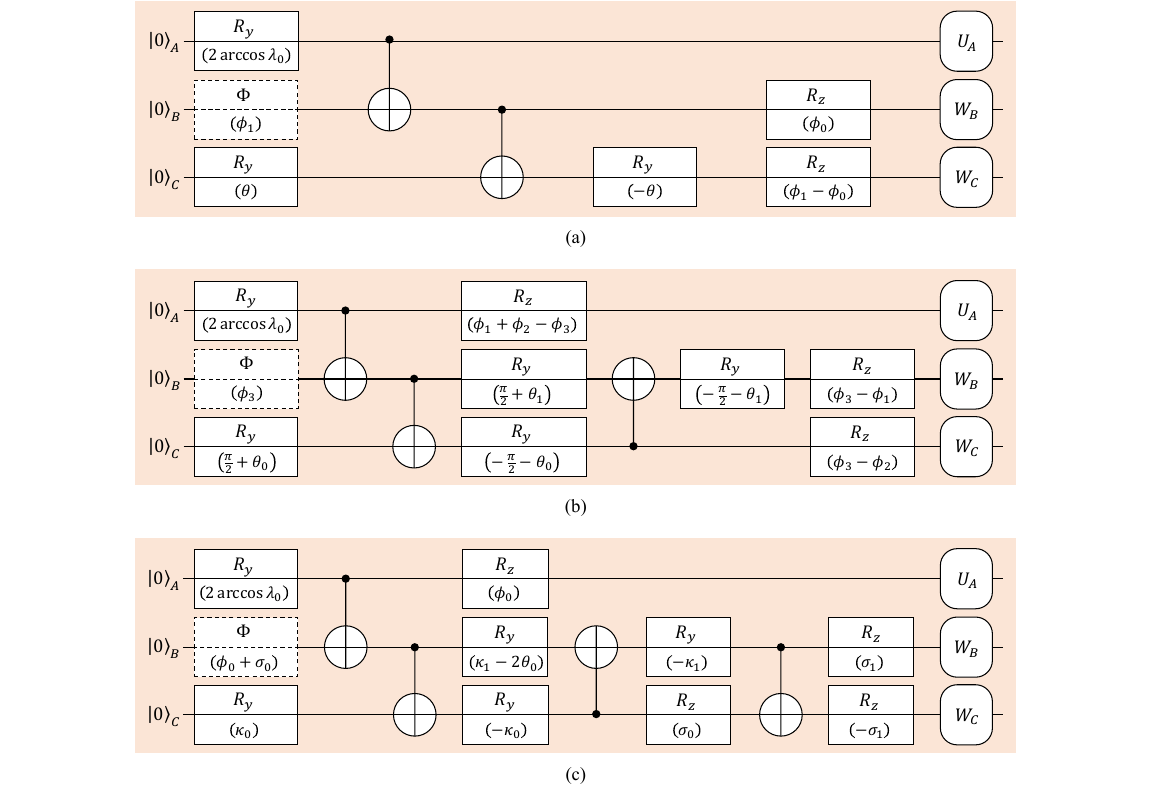}
\caption{Circuits for preparing an $A|BC$-entangled three-qubit state $\ket{\psi_3}$. Based on the Schmidt decomposition in Eq.~(\ref{eq:Psi3SD}), we construct state-preparation circuits for the SS-, SE-, and EE-type families.
(a) Circuit for preparing SS-type states. The angles $\theta$ and $\phi_j$ are given in Eq.~(\ref{eq:Mapping3}), and the final single-qubit unitaries $U_A$, $W_B$, and $W_C$ are defined in Eqs.~(\ref{eq:Ua}) and~(\ref{eq:WbWc}). 
(b) Circuit for preparing SE-type states. The circuit parameters $\theta_j$ and $\phi_j$ are specified in Eq.~(\ref{eq:SEVector}), and $U_A$, $W_B$, and $W_C$ are defined in Eqs.~(\ref{eq:Ua}) and~(\ref{eq:SEWbSEWc}). 
(c) Circuit for preparing EE-type states. The phase $\phi_0$ and the parameters $\kappa_j$ and $\sigma_j$ are specified in Eq.~(\ref{eq:EEVector}) and in Eqs.~(\ref{eq:Kappa0})--(\ref{eq:Sigma1}), respectively. The final single-qubit unitaries $U_A$, $W_B$, and $W_C$ are defined in Eqs.~(\ref{eq:Ua}) and~(\ref{eq:EEWbWc}). 
}
\label{fig:SP3}
\end{figure*}

We first consider the case in which the qubit states on subsystem $B$ are orthogonal, i.e., $\inn{\gamma_0}{\gamma_1}=0$. In this situation, while preserving both the orthogonality and the separability of the Schmidt basis vectors, we can transform the target state into the form
\begin{equation}
U_A^\dagger \otimes W_B^\dagger \otimes W_C^\dagger \ket{\psi_3}_{ABC}
=
\lambda_0 \ket{000}_{ABC}
+
\lambda_1 \ket{1}_A \otimes \ket{1}_B \otimes W_C^\dagger\ket{\delta_1}_C,
\end{equation}
where the single-qubit unitary $U_A$ is given in Eq.~(\ref{eq:Ua}), and the remaining unitaries $W_B$ and $W_C$ are defined by
\begin{equation}
W_B \coloneq \ket{\gamma_0}\bra{0} + \ket{\gamma_1}\bra{1}
\quad \mathrm{and} \quad
W_C \coloneq \ket{\delta_0}\bra{0} + \ket{\delta_0^\perp}\bra{1}. \label{eq:WbWc}
\end{equation}
This representation makes it clear what core transformation must be realized by the two-qubit unitary $U_{BC}$ for SS-type states. Up to single-qubit unitaries applied before and after $U_{BC}$, it suffices to implement the mapping
\begin{eqnarray}
\ket{0} \otimes \ket{0} &\mapsto& \ket{0} \otimes \ket{0}, \\
\ket{1} \otimes \ket{0} &\mapsto& \ket{1} \otimes \bigl(\sin \theta\, e^{i \phi_0} \ket{0} + \cos \theta\, e^{i \phi_1} \ket{1}\bigr), \label{eq:Mapping3}
\end{eqnarray}
where $\theta\in[0,\pi/2]$ and $\phi_j\in\mathbb{R}$ for $j=0,1$. Since the two input states are orthogonal in the first qubit, this mapping can be implemented using a single CNOT gate. Concretely, the required conditional action on the second qubit is realized by the standard construction of Ref.~\cite{Barenco1995}:
\begin{equation}
R_z(\phi_1-\phi_0)R_y(-\theta)XR_y(\theta)R_z(\phi_0-\phi_1)
=
\begin{bmatrix}
\sin \theta & \cos \theta e^{i (\phi_0-\phi_1)} \\
\cos \theta e^{i (\phi_1-\phi_0)} & -\sin \theta
\end{bmatrix},
\end{equation}
where the Pauli-$X$ gate is
\begin{equation}
X = \begin{bmatrix} 0 & 1 \\ 1 & 0 \end{bmatrix}.
\end{equation}
When combined with a CNOT, this construction acts nontrivially only when the control qubit is in the state $\ket{1}$, i.e., on the $\ket{10}$ component. The resulting circuit is shown in Fig.~\ref{fig:SP3}(a).

Next, we consider the case in Eq.~(\ref{eq:SS}) in which the qubit states on subsystem $C$ are orthogonal, i.e., $\inn{\delta_0}{\delta_1}=0$. This case is equivalent to the previous one, since it is obtained by interchanging the roles of systems $B$ and $C$. Operationally, the corresponding state-preparation circuit is obtained by swapping the target qubit of the gates in Fig.~\ref{fig:SP3}(a).

\subsection{State-preparation circuit for SE-type states} \label{sec:SE}

We present an explicit circuit for preparing an SE-type target state $\ket{\psi_3}$. In analogy with the SS-type case, we identify the core action that the two-qubit unitary $U_{BC}$ in Fig.~\ref{fig:SP3General} must realize. Assume that the first Schmidt basis vector $\ket{\beta_{00}}$ is separable and the second one $\ket{\beta_{10}}$ is entangled. Specifically, we write
\begin{equation}
\ket{\beta_{00}} = \ket{\gamma_0} \otimes \ket{\delta_0},
\end{equation}
where $\ket{\gamma_0}$ and $\ket{\delta_0}$ are arbitrary single-qubit pure states. By applying suitable single-qubit unitaries to each subsystem, we can transform the target state $\ket{\psi_3}$ while preserving its entanglement structure and obtain
\begin{equation}
U_A^\dagger \otimes W_B^\dagger \otimes W_C^\dagger \ket{\psi_3}_{ABC}
=
\lambda_0 \ket{000}_{ABC}
+
\lambda_1 \ket{1}_A \otimes ( W_B^\dagger \otimes W_C^\dagger ) \ket{\beta_{10}}_{BC},
\end{equation}
where the single-qubit unitary $U_A$ is given in Eq.~(\ref{eq:Ua}), and the remaining unitaries $W_B$ and $W_C$ are defined as
\begin{equation}
W_B \coloneq \ket{\gamma_0}\bra{0} + \ket{\gamma_0^\perp}\bra{1}
\quad \mathrm{and} \quad
W_C \coloneq \ket{\delta_0}\bra{0} + \ket{\delta_0^\perp}\bra{1}. \label{eq:SEWbSEWc}
\end{equation}
Proceeding exactly as in the SS-type case, we find that the essential mapping to be implemented for SE-type states is
\begin{eqnarray}
\ket{0} \otimes \ket{0} &\mapsto& \ket{0} \otimes \ket{0}, \\
\ket{1} \otimes \ket{0} &\mapsto&
\sin\theta_0\sin\theta_1 e^{i\phi_1} \ket{0} \otimes \ket{1} \,
+ \cos\theta_0 e^{i\phi_2} \ket{1} \otimes \ket{0}
+ \sin\theta_0\cos\theta_1 e^{i\phi_3} \ket{1} \otimes \ket{1}, \label{eq:SEVector}
\end{eqnarray}
where $\theta_0\in(0,\pi/2)$, $\theta_1\in(0,\pi/2]$, and $\phi_j\in\mathbb{R}$ for $j=1,2,3$. By varying $\theta_0$, $\theta_1$, and $\phi_j$, the right-hand side ranges over all normalized two-qubit states orthogonal to $\ket{00}$.

We now explain how to implement the above mapping. Ignoring the relative phases $e^{i\phi_j}$, the magnitudes of the three nonzero amplitudes can be set using two $R_y$ rotations. Since $\ket{00}$ must remain unchanged, these rotations must be implemented conditionally. Using the standard construction of Ref.~\cite{Barenco1995}, a controlled rotation on the target qubit can be realized by inserting a CNOT gate, with this qubit as target, between two single-qubit rotations on the same qubit, which yields
\begin{equation}
R_y(-\theta)XR_y(\theta)
= \begin{bmatrix}
\sin \theta & \cos \theta \\
\cos \theta & -\sin \theta
\end{bmatrix}
\quad \mathrm{and} \quad
R_y(-\theta)IR_y(\theta) = I.
\end{equation}
Thus, when the control qubit is in the state $\ket{1}$, the target undergoes the desired nontrivial transformation, whereas when the control qubit is in the state $\ket{0}$ the overall action on the target is the identity. After fixing the magnitudes in this manner, we place $R_z$ rotations before and after the $R_y$ rotations to incorporate the phase parameters $\phi_j$. Figure~\ref{fig:SP3}(b) depicts the resulting circuit for preparing SE-type states.

Finally, consider the complementary situation in which $\ket{\beta_{00}}$ is entangled and $\ket{\beta_{10}}$ is separable. This case is equivalent to the one treated above: by interchanging the two Schmidt terms, that is, by swapping the Schmidt coefficients and the corresponding Schmidt basis vectors, the same circuit template can be used without further modification.

\subsection{State-preparation circuit for EE-type states} \label{sec:EE}

Finally, we present a concrete circuit for preparing EE-type states. As in the SS- and SE-type cases, the central step is to isolate the essential action of the two-qubit unitary $U_{BC}$ in Fig.~\ref{fig:SP3General}. Suppose that the first Schmidt basis vector admits the Schmidt decomposition
\begin{equation}
\ket{\beta_{00}}
= \cos \theta_0 \ket{\gamma_0} \otimes \ket{\delta_0}
+\sin \theta_0 \left(-\ket{\gamma_1}\right) \otimes \ket{\delta_1},
\end{equation}
where $\theta_0\in(0,\pi/2)$, and $\{(-1)^j\ket{\gamma_j}\}_{j=0,1}$ and $\{\ket{\delta_j}\}_{j=0,1}$ form orthonormal bases for the single-qubit subsystems $B$ and $C$, respectively.

To make the required structure explicit, we apply suitable single-qubit unitaries and rewrite the target state as
\begin{equation}
U_A^\dagger \otimes W_B^\dagger \otimes W_C^\dagger \ket{\psi_3}_{ABC}
=
\lambda_0 \ket{0}_A \otimes \left( \cos \theta_0 \ket{00}_{BC} - \sin \theta_0 \ket{11}_{BC} \right)
+ \lambda_1 \ket{1}_A \otimes ( W_B^\dagger \otimes W_C^\dagger ) \ket{\beta_{10}}_{BC},
\end{equation}
where the single-qubit unitary $U_A$ is given in Eq.~(\ref{eq:Ua}), and the unitaries $W_B$ and $W_C$ are defined by
\begin{equation}
W_B \coloneq \sum_{j=0}^1 \ket{\gamma_j}\bra{j}
\quad \mathrm{and} \quad
W_C \coloneq \sum_{j=0}^1 \ket{\delta_j}\bra{j}. \label{eq:EEWbWc}
\end{equation}
In this form, the core mapping that $U_{BC}$ must realize is
\begin{eqnarray}
\ket{0} \otimes \ket{0} &\mapsto&
\cos \theta_0 \ket{0} \otimes \ket{0} - \sin \theta_0 \ket{1} \otimes \ket{1}, \\
\ket{1} \otimes \ket{0} &\mapsto&
 \cos\theta_1\sin\theta_0 e^{i\phi_0}\ket{0} \otimes \ket{0}
+ \sin\theta_1\sin\theta_2 e^{i\phi_1}\ket{0} \otimes \ket{1} \nonumber\\
&&+ \sin\theta_1\cos\theta_2 e^{i\phi_2}\ket{1} \otimes \ket{0}
+ \cos\theta_1\cos\theta_0 e^{i\phi_0}\ket{1} \otimes \ket{1}, \label{eq:EEVector}
\end{eqnarray}
where $\theta_1,\theta_2\in[0,\pi/2]$ and $\phi_j\in\mathbb{R}$ for $j=0,1,2$. By varying $\theta_1$, $\theta_2$, and $\phi_j$, the image of $\ket{1}\otimes\ket{0}$ ranges over all normalized two-qubit states orthogonal to $\cos \theta_0 \ket{00} - \sin \theta_0 \ket{11}$.

The above mapping can be realized by taking the SE-type core circuit in Fig.~\ref{fig:SP3}(b) as a starting point and extending it by one additional relative phase gate and one extra CNOT gate, with the gate parameters chosen appropriately. For this purpose, we introduce the parameters
\begin{align}
\kappa_0 &\coloneqq \frac{\pi}{2} + \theta_1, \quad 
\kappa_1 \coloneqq \frac{\pi}{2} + \theta_0 - \theta_2, \label{eq:Kappa0} \\
\sigma_0 &\coloneqq \frac{\phi_1 + \phi_2 - \pi}{2} - \phi_0, \quad
\sigma_1 \coloneqq \frac{\pi - \phi_1 + \phi_2}{2}, \label{eq:Sigma1}
\end{align}
which parametrize the gates in the EE-type circuit. The complete circuit for preparing this class of EE-type $A|BC$-entangled states is shown in Fig.~\ref{fig:SP3}(c).

\section{Four Classes of Three-Qubit States with Explicit Preparation Circuits} \label{sec:Examples}

Our state-preparation circuits enable the systematic preparation of an arbitrary three-qubit pure state. As an application of our general construction, we present explicit state-preparation circuits for a collection of three-qubit pure states~\cite{Greenberger1990, Duer2000, Fortescue2007, Briegel2001, Singh2023, Hein2004, Roy2015, Eltschka2008, Sabin2008, Li2014, LunaHernandez2024, Man2007, Sun2012, Nie2011, Artawan2019, Marconi2025, Frydryszak2009, Prakash2011, Yang2009, Zhu2014, Acin2001, Singh2022, Aulbach2010, Nusur2021, Huang2021, Huang2024} that are widely used in quantum information theory and related areas of quantum science.

To illustrate the applicability of our circuits, we group these states into four classes, denoted by $\mathrm{R}_k$. Here $\mathrm{R}_k$ denotes the $k$th class, defined in the $k$th subsection as a superposition over a fixed set of computational-basis states, and we provide explicit preparation circuits for each class. A key advantage of the resulting circuits is that they are parametrized directly by the amplitude and phase data of $\mathrm{R}_k$-class states. Consequently, by adjusting only the gate parameters, one can readily prepare an arbitrary target state within the $\mathrm{R}_k$ class. Furthermore, any state that is locally unitarily equivalent to a given $\mathrm{R}_k$-class state can be prepared by appending appropriate single-qubit gates to the outputs of the circuit. This demonstrates that our scheme yields practical, directly implementable circuits for preparing known three-qubit states.

\subsection{$\mathrm{R}_1$ class: Superpositions of $\ket{000}$ and $\ket{111}$}

As a first application of our results, we focus on the three-qubit GHZ state~\cite{Greenberger1990,Duer2000} and several of its variations~\cite{Fortescue2007,Briegel2001,Singh2023,Hein2004,Roy2015,Eltschka2008}. These states serve as standard examples of genuine tripartite entanglement and nonlocal correlations~\cite{Greenberger1990,Duer2000}, and they play a central role in a variety of tasks in multipartite quantum communication~\cite{Fortescue2007,Singh2023} and measurement-based quantum computation~\cite{Briegel2001,Hein2004}.

A common feature of these three-qubit pure states is that they can all be written as a superposition of the computational-basis states $\ket{000}$ and $\ket{111}$. In particular, they belong to the $\mathrm{R}_1$ class, consisting of states of the form
\begin{equation} \label{eq:R1}
\ket{\psi_{\mathrm{R}_1}} = r_0 e^{i\varphi_0} \ket{000} + r_1 e^{i\varphi_1} \ket{111},
\end{equation}
where $r_0,r_1 > 0$ with $r_0^2 + r_1^2 = 1$ and $\varphi_0,\varphi_1 \in \mathbb{R}$. Compared with other three-qubit pure states, the $\mathrm{R}_1$ class has a particularly simple entanglement structure. Its Schmidt decomposition with respect to the $A|BC$ bipartition is given by
\begin{equation}
\ket{\psi_{\mathrm{R}_1}}
= r_0 \ket{0} \otimes \left( e^{i\varphi_0} \ket{00} \right)
+ r_1 \ket{1} \otimes \left( e^{i\varphi_1} \ket{11} \right).
\end{equation}
According to our entanglement classification, the state $\ket{\psi_{\mathrm{R}_1}}$ is thus SS-type. To obtain a circuit that prepares this class, we determine the gate parameters and single-qubit unitaries in the SS-type template shown in Fig.~\ref{fig:SP3}(a).

\begin{figure*}
\centering
\includegraphics[width=\textwidth]{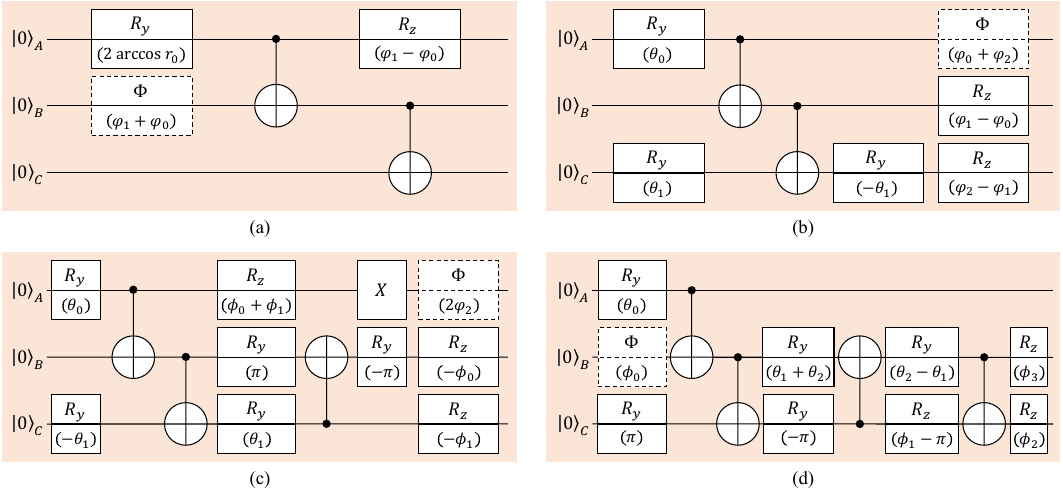}
\caption{Explicit circuits for preparing the $\mathrm{R}_1$--$\mathrm{R}_4$ classes.
(a) Circuit for the $\mathrm{R}_1$ class defined in Eq.~(\ref{eq:R1}).
(b) Circuit for the $\mathrm{R}_2$ class. The phases $\varphi_j$ appear in Eq.~(\ref{eq:R2}), and the rotation angles $\theta_j$ are given in Eq.~(\ref{eq:R2Theta}).
(c) Circuit for the $\mathrm{R}_3$ class. The gate parameters are determined by the state parameter $\varphi_2$ in Eq.~(\ref{eq:R3}) and by $\theta_j$ and $\phi_j$ in Eq.~(\ref{eq:R3Phi}).
(d) Circuit for the $\mathrm{R}_4$ class. The state parameters $r_j$ and $\varphi_j$ in Eq.~(\ref{eq:R4}) determine the gate parameters $\theta_j$ and $\phi_j$ in Eqs.~(\ref{eq:R4Theta0})--(\ref{eq:R4Phi2}).
By choosing appropriate parameter values, these circuits prepare the states listed in Table~\ref{tab:All}.}
\label{fig:All}
\end{figure*}

The resulting circuit is depicted in Fig.~\ref{fig:All}(a) and consists of gates whose parameters are expressed solely in terms of the amplitudes $r_j$ and phases $\varphi_j$ of the target state $\ket{\psi_{\mathrm{R}_1}}$. From an implementation perspective, the global phase gate $\Phi$ in the circuit can be omitted. Moreover, the relative phase gate $P$ may be applied to either the first or the second qubit without changing the output state up to a global phase. In particular, if $P$ is moved to the first qubit, the target state can be prepared with a depth-3 circuit. Several representative $\mathrm{R}_1$-class states are listed in Table~\ref{tab:All}, together with their computational-basis coefficients.

\begin{table}
\caption{Representative examples of three-qubit pure states belonging to the $\mathrm{R}_1$--$\mathrm{R}_4$ classes. The first column indicates the class membership. The next eight columns list the coefficients associated with the computational-basis states. The tenth column specifies any additional local unitaries (LUs) applied after preparing the target state, and the last column lists the state names used in the cited references. Here an en dash ``--'' indicates a zero coefficient in Columns~2--9 and indicates that no LUs are applied in Column~10. Here $p>0$ and $p_j>0$, $q_j\ge 0$ with $q_0 \ge q_1 \ge q_2$, $n$ is a positive integer, and $H$ denotes the Hadamard gate~\cite{Nielsen2010,Wilde2013}.}
\label{tab:All}
\centering
\begin{tabular}{ccccccccccl}
\hline
Class & $\ket{000}$ & $\ket{001}$ & $\ket{010}$ & $\ket{011}$ & $\ket{100}$ & $\ket{101}$ & $\ket{110}$ & $\ket{111}$ & LUs & Name \\
\hline\hline
\multirow{4}{*}[-1.0ex]{$\mathrm{R}_1$} & $\frac{1}{\sqrt{2}}$ & -- & -- & -- & -- & -- & -- & $\frac{1}{\sqrt{2}}$ & -- & GHZ state~\cite{Duer2000} \\
 & $\frac{1}{\sqrt{1+p^2}}$ & -- & -- & -- & -- & -- & -- & $\frac{p}{\sqrt{1+p^2}}$ & -- & GHZ-type state~\cite{Roy2015} \\
 & $r_0 e^{i\varphi_0}$ & -- & -- & -- & -- & -- & -- & $r_1 e^{i\varphi_1}$ & -- & GHZ-like state~\cite{Fortescue2007}, Generalized GHZ state~\cite{Eltschka2008} \\
 & $\frac{1}{\sqrt{2}}$ & -- & -- & -- & -- & -- & -- & $\frac{1}{\sqrt{2}}$ & $H\otimes I\otimes H$ & Cluster state~\cite{Briegel2001,Singh2023}, Graph state $\ket{P_3}$~\cite{Hein2004} \\
\hline
\multirow{3}{*}[-0.5ex]{$\mathrm{R}_2$} & $\frac{1}{\sqrt{2}}$ & -- & -- & -- & -- & -- & $\frac{r_0}{\sqrt{2}}$ & $\frac{r_1}{\sqrt{2}}$ & -- & Maximal slice state~\cite{Li2014} \\
 & $r_0 e^{i\varphi_0}$ & -- & -- & -- & -- & -- & $r_1 e^{i\varphi_1}$ & $r_2 e^{i\varphi_2}$ & -- & Type $IV''$ state~\cite{Sabin2008} \\
 & $p_0$ & -- & -- & -- & -- & -- & $p_1$ & $p_2$ & -- & Type 3b-3 state~\cite{LunaHernandez2024} \\
\hline
\multirow{6}{*}[-1.0ex]{$\mathrm{R}_3$} & -- & $\frac{1}{\sqrt{3}}$ & $\frac{1}{\sqrt{3}}$ & -- & $\frac{1}{\sqrt{3}}$ & -- & -- & -- & -- & W state~\cite{Duer2000}, Dicke state $\ket{D_1^3}$~\cite{Marconi2025} \\
 & -- & $\frac{1}{\sqrt{2}}$ & $\frac{\sqrt{n}}{\sqrt{2n+2}}$ & -- & $\frac{1}{\sqrt{2n+2}}$ & -- & -- & -- & -- & $\mathrm{W}_n$ state~\cite{Roy2015} \\
 & -- & $q_0$ & $q_1$ & -- & $q_2$ & -- & -- & -- & -- & W-like state~\cite{Fortescue2007} \\
 & -- & $\frac{1}{\sqrt{2}}$ & $\frac{1}{2}$ & -- & $\frac{1}{2}$ & -- & -- & -- & -- & W-class state~\cite{Nie2011,Artawan2019} \\
 & -- & $r_0 e^{i\varphi_0}$ & $r_1 e^{i\varphi_1}$ & -- & $r_2 e^{i\varphi_2}$ & -- & -- & -- & -- & Generalized W state~\cite{Eltschka2008}, W-like state~\cite{Man2007,Sun2012} \\
 & -- & $\frac{1}{\sqrt{3}}$ & $\frac{1}{\sqrt{3}}$ & -- & $\frac{1}{\sqrt{3}}$ & -- & -- & -- & $X^{\otimes 3}$ & Dicke state $\ket{D_2^3}$~\cite{Marconi2025}, Cluster Werner state~\cite{Frydryszak2009} \\
\hline
\multirow{6}{*}[-1.0ex]{$\mathrm{R}_4$} & $\frac{1}{2}$ & -- & -- & $\pm\frac{1}{2}$ & -- & $\pm\frac{1}{2}$ & $\pm\frac{1}{2}$& -- & -- & GHZ-like state~\cite{Prakash2011,Yang2009,Zhu2014} \\
 & $-\frac{1}{2}$ & -- & -- & $\frac{1}{2}$ & -- & $\frac{1}{2}$ & $\frac{1}{2}$ & -- & $(HX)^{\otimes 3}$ & Graph state $\ket{K_3}$~\cite{Hein2004} \\
 & $\frac{1}{2}$ & -- & -- & $\frac{1}{2}$ & -- & $\frac{1}{2}$ & $\frac{1}{2}$ & -- & -- & Phase-flip code~\cite{Huang2021,Huang2024} \\
 & $\frac{1}{2}$ & -- & -- & $\frac{1}{2}$ & -- & $\frac{1}{2}$ & $\frac{1}{2}$ & -- & $X^{\otimes 3}$ & Phase-flip code~\cite{Huang2021,Huang2024} \\
 & $\frac{\sqrt{3}}{2}$ & -- & -- & $\frac{1}{2\sqrt{3}}$ & -- & $\frac{1}{2\sqrt{3}}$ & $\frac{1}{2\sqrt{3}}$ & -- & $X^{\otimes 3}$ & Decay state~\cite{Acin2001} \\
 & $\frac{1}{\sqrt{2}}$ & -- & -- & $\frac{-1}{\sqrt{6}}$ & -- & $\frac{1}{\sqrt{6}}$ & $\frac{1}{\sqrt{6}}$ & -- & $X^{\otimes 3}$ & Permutation-symmetric state~\cite{Singh2022,Aulbach2010} \\
\hline
\end{tabular}
\end{table}

\subsection{$\mathrm{R}_2$ class: Superpositions of $\ket{000}$, $\ket{110}$, $\ket{111}$}

As a second application, we consider another SS-type family, which we denote by the $\mathrm{R}_2$ class. These states play an important role in three-qubit entanglement classification~\cite{Sabin2008}, controlled teleportation protocols~\cite{Li2014}, and the analysis of multipartite entanglement measures~\cite{LunaHernandez2024}.

In Refs.~\cite{Sabin2008,Li2014,LunaHernandez2024}, the corresponding three-qubit states can be written as a superposition of $\ket{000}$, $\ket{110}$, and $\ket{111}$. We therefore consider the most general superposition of this form,
\begin{equation} \label{eq:R2}
\ket{\psi_{\mathrm{R}_2}}
= r_0 e^{i\varphi_0} \ket{000}
+ r_1 e^{i\varphi_1} \ket{110}
+ r_2 e^{i\varphi_2} \ket{111},
\end{equation}
where $r_j>0$ for $j=0,1,2$ with $\sum_{j=0}^2 r_j^2 = 1$ and $\varphi_j\in\mathbb{R}$. Compared with generic three-qubit pure states, the $\mathrm{R}_2$ class also has a simple entanglement structure with respect to the $A|BC$ bipartition. The state $\ket{\psi_{\mathrm{R}_2}}$ admits the Schmidt decomposition
\begin{equation}
\ket{\psi_{\mathrm{R}_2}}
 = r_0 \ket{0} \otimes \ket{\beta_{00}}
 + \sqrt{r_1^2 + r_2^2} \ket{1} \otimes \ket{\beta_{10}},
\end{equation}
where the Schmidt basis states $\ket{\beta_{j0}}$ are given by
\begin{equation}
\ket{\beta_{00}} = e^{i\varphi_0} \ket{00} \quad \mathrm{and} \quad
\ket{\beta_{10}} = \frac{r_1}{\sqrt{r_1^2 + r_2^2}} e^{i\varphi_1} \ket{10}
 + \frac{r_2}{\sqrt{r_1^2 + r_2^2}} e^{i\varphi_2} \ket{11}.
\end{equation}
According to our entanglement classification, the target state $\ket{\psi_{\mathrm{R}_2}}$ is therefore an SS-type state. Based on this, we obtain an explicit circuit in Fig.~\ref{fig:All}(b) that prepares $\ket{\psi_{\mathrm{R}_2}}$. By instantiating the SS-type template in Fig.~\ref{fig:SP3}(a), we arrive at a circuit whose rotation angles are given by
\begin{equation} \label{eq:R2Theta}
\theta_0 = 2 \arccos r_0 \quad \mathrm{and} \quad \theta_1 = \arccos \left( \frac{r_2}{\sqrt{r_1^2 + r_2^2}} \right).
\end{equation}

Representative $\mathrm{R}_2$-class states are listed in Table~\ref{tab:All}, together with the corresponding coefficients of the computational-basis states. In particular, with the parameter choices listed in Table~\ref{tab:All}, the circuit in Fig.~\ref{fig:All}(b) can be used directly as a state-preparation module for the type $IV''$, maximal slice, and 3b-3 states appearing in Refs.~\cite{Sabin2008,Li2014,LunaHernandez2024}.

\subsection{$\mathrm{R}_3$ class: Superpositions of $\ket{001}$, $\ket{010}$, $\ket{100}$}

As another application of our construction, we consider the three-qubit W state~\cite{Duer2000} and several of its variations~\cite{Eltschka2008,Man2007,Sun2012,Roy2015,Fortescue2007,Nie2011,Artawan2019,Marconi2025,Frydryszak2009}. Together with the GHZ state, the W state is widely regarded as a canonical genuinely entangled three-qubit state and has been extensively studied as a resource for various tasks in quantum information processing.

All of these states lie in the single-excitation subspace spanned by the computational-basis states $\ket{001}$, $\ket{010}$, and $\ket{100}$. Motivated by this observation, we introduce the $\mathrm{R}_3$ class defined as
\begin{equation} \label{eq:R3}
\ket{\psi_{\mathrm{R}_3}}
 = r_0 e^{i\varphi_0} \ket{001}
 + r_1 e^{i\varphi_1} \ket{010}
 + r_2 e^{i\varphi_2} \ket{100},
\end{equation}
where $r_j>0$ for $j=0,1,2$ with $\sum_{j=0}^2 r_j^2 = 1$ and $\varphi_j\in\mathbb{R}$. With respect to the $A|BC$ bipartition, the target state $\ket{\psi_{\mathrm{R}_3}}$ has the Schmidt decomposition
\begin{equation}
\ket{\psi_{\mathrm{R}_3}}
 = r_2 \ket{1} \otimes \ket{\beta_{00}}
 + \sqrt{r_0^2+r_1^2} \ket{0} \otimes \ket{\beta_{10}},
\end{equation}
where the Schmidt basis states $\ket{\beta_{j0}}$ are given by
\begin{equation}
\ket{\beta_{00}} = e^{i\varphi_2} \ket{00} \quad \mathrm{and} \quad
\ket{\beta_{10}} = \frac{r_0}{\sqrt{r_0^2+r_1^2}} e^{i\varphi_0} \ket{01}
 + \frac{r_1}{\sqrt{r_0^2+r_1^2}} e^{i\varphi_1} \ket{10}.
\end{equation}
This Schmidt decomposition implies that $\ket{\psi_{\mathrm{R}_3}}$ belongs to the SE type in our classification. Consequently, a circuit that prepares the target state can be obtained by specializing the SE-type template shown in Fig.~\ref{fig:SP3}(b). In this case, the gate parameters are chosen as
\begin{equation}
\theta_0 = 2 \arccos r_2, \quad
\theta_1 = - \frac{\pi}{2} - \arccos \left( \frac{r_1}{\sqrt{r_0^2 + r_1^2}} \right), \quad \mathrm{and} \quad
\phi_j = \varphi_j - \varphi_2, \label{eq:R3Phi}
\end{equation}
where $\phi_j$ is defined for $j=0,1,2$. The resulting circuit implementation is depicted in Fig.~\ref{fig:All}(c). Representative $\mathrm{R}_3$-class states are listed in Table~\ref{tab:All}.

\subsection{$\mathrm{R}_4$ class: Superpositions of $\ket{000}$, $\ket{011}$, $\ket{101}$, $\ket{110}$}

We now turn to a special subclass of EE-type three-qubit states. Unlike the previous classes, this family is supported on the four computational-basis states $\ket{000}$, $\ket{011}$, $\ket{101}$, and $\ket{110}$. Superpositions over these basis states arise in a variety of settings, including multipartite graph states~\cite{Hein2004}, teleportation and controlled-teleportation schemes~\cite{Prakash2011,Yang2009}, and the study of multipartite entanglement~\cite{Zhu2014,Acin2001,Singh2022,Aulbach2010} and nonlocality~\cite{Nusur2021,Huang2021,Huang2024}.

In practical implementations of these protocols, one typically requires an efficient method to prepare such states as an initial step. With this motivation, we define the $\mathrm{R}_4$ class as
\begin{equation} \label{eq:R4}
\ket{\psi_{\mathrm{R}_4}} = r_0 e^{i\varphi_0} \ket{000} + r_1 e^{i\varphi_1} \ket{011} + r_2 e^{i\varphi_2} \ket{101} + r_3 e^{i\varphi_3} \ket{110},
\end{equation}
where $r_j>0$ for $j=0,1,2,3$ with $\sum_{j=0}^3 r_j^2 = 1$ and $\varphi_j\in\mathbb{R}$. By appropriate choices of these parameters, this class captures a range of notable three-qubit states. With respect to the $A|BC$ bipartition, the target state $\ket{\psi_{\mathrm{R}_4}}$ admits the Schmidt decomposition
\begin{equation}
\ket{\psi_{\mathrm{R}_4}}
 = \sqrt{r_0^2+r_1^2} \ket{0} \otimes \ket{\beta_{00}}
 + \sqrt{r_2^2+r_3^2} \ket{1} \otimes \ket{\beta_{10}},
\end{equation}
where the Schmidt basis states $\ket{\beta_{j0}}$ are given by
\begin{equation}
\ket{\beta_{00}} = \frac{r_0}{\sqrt{r_0^2+r_1^2}} e^{i\varphi_0} \ket{00}
 + \frac{r_1}{\sqrt{r_0^2+r_1^2}} e^{i\varphi_1} \ket{11} \quad \mathrm{and} \quad
\ket{\beta_{10}} = \frac{r_2}{\sqrt{r_2^2+r_3^2}} e^{i\varphi_2} \ket{01}
 + \frac{r_3}{\sqrt{r_2^2+r_3^2}} e^{i\varphi_3} \ket{10}.
\end{equation}
This decomposition leads directly to an explicit preparation circuit. Concretely, specializing the EE-type template in Fig.~\ref{fig:SP3}(c) yields the circuit shown in Fig.~\ref{fig:All}(d). For a compact specification of this circuit, we introduce gate parameters $\theta_j$ and $\phi_j$ in terms of the state parameters $r_j$ and $\varphi_j$ as
\begin{align}
\theta_0 &= 2 \arccos \sqrt{r_0^2 + r_1^2}, \quad
\theta_1 = \frac{\pi}{2} - \arccos \left( \frac{r_3}{\sqrt{r_2^2 + r_3^2}} \right), \quad
\theta_2 = - \arccos \left( \frac{r_0}{\sqrt{r_0^2 + r_1^2}} \right), \label{eq:R4Theta0} \\
\phi_0 &= \frac{+\varphi_0+\varphi_1+\varphi_2+\varphi_3}{2}, \quad
\phi_1 = \frac{-\varphi_0-\varphi_1+\varphi_2+\varphi_3}{2}, \\
\phi_2 &= \frac{-\varphi_0+\varphi_1+\varphi_2-\varphi_3}{2}, \quad
\phi_3 = \frac{-\varphi_0+\varphi_1-\varphi_2+\varphi_3}{2} + \pi. \label{eq:R4Phi2}
\end{align}
Representative examples within the $\mathrm{R}_4$ class are listed in Table~\ref{tab:All}.

In summary, this section demonstrated how our general state-preparation framework can be specialized to efficiently generate a broad set of representative three-qubit pure states appearing in the quantum-information literature. By grouping these states into four classes $\mathrm{R}_k$ and deriving explicit preparation circuits for each class, we obtained implementations whose gate parameters are determined directly by the amplitude and phase data of the target state, so that arbitrary $\mathrm{R}_k$-class states can be prepared simply by tuning these parameters; moreover, states that are locally unitarily equivalent to a given $\mathrm{R}_k$-class state can be obtained by appending suitable single-qubit gates.

\begin{table}[t]
\centering
\caption{Resource comparison between the fixed-depth template of \cite{Giraud2009} and the circuits obtained for each $\mathrm{R}_k$ class in this work, assuming that CNOT gates are available only between adjacent qubits.}
\label{tab:resource_comparison}
\begin{tabular}{lccccc}
\hline
Circuit / class & $n(R_z)$ & $n(R_y)$ & $n(\mathrm{CNOT})$ & Total & Depth \\
\hline
Fixed template~\cite{Giraud2009} & 7 & 8 & 6 & 21 & 14 \\
$\mathrm{R}_1$ class & 1 & 1 & 2 & 4 & 3 \\
$\mathrm{R}_2$ class & 2 & 3 & 2 & 7 & 5 \\
$\mathrm{R}_3$ class & 3 & 5 & 3 & 11 & 7 \\
$\mathrm{R}_4$ class & 3 & 5 & 4 & 12 & 8 \\
\hline
\end{tabular}
\end{table}

From an implementation perspective, exploiting the restricted support and entanglement structure of each class leads to substantial resource savings compared with a fixed-depth template. As shown in Table~\ref{tab:resource_comparison}, while the fixed template of Ref.~\cite{Giraud2009} requires 18 gates at depth 11, our class-adapted circuits reduce the total gate count to 4--12 and the depth to 3--8 for the $\mathrm{R}_1$--$\mathrm{R}_4$ classes. These reductions are particularly pronounced for the simpler classes $\mathrm{R}_1$ and $\mathrm{R}_2$, while remaining significant even for the most general case $\mathrm{R}_4$.

\section{Conclusion} \label{sec:Conclusion}

In this work, we developed a fully explicit approach to deterministic three-qubit state preparation based on the Schmidt decomposition with respect to the $A|BC$ bipartition. Starting from an arbitrary target state specified by its computational-basis amplitudes, we presented a concrete procedure to (i) identify its entanglement type, (ii) extract the associated structural data, including Schmidt coefficients and Schmidt bases, and (iii) map this data to a state-preparation circuit via a small set of canonical circuit modules. Along the way, we provided explicit circuits for each entanglement type and for several practically relevant subclasses, thereby yielding a structured compilation framework for three-qubit state preparation.

A key feature of our method is that it is constructive and parameter-explicit: once the target amplitudes are given, the required gate parameters follow from a deterministic sequence of analytic steps~\cite{Moettoenen2005,Plesch2011}. In particular, the synthesis does not depend on ad hoc local tweaks, implicit gauge choices, or case-by-case hidden decompositions; instead, our pipeline has a well-defined input--output specification at every stage, including entanglement-type classification, Schmidt-data extraction, and circuit instantiation.

Our circuits are also connectivity-aware in a practically relevant sense. On NISQ-era superconducting platforms such as IBM Quantum devices, the available two-qubit interactions are constrained by a sparse coupling map~\cite{IBMHeavyHex2021}, and naive decompositions of three-qubit unitaries can easily incur additional routing overhead, most notably SWAP insertions~\cite{Li2019}. By organizing the preparation around the $A|BC$ split and arranging the two-qubit interactions accordingly, our constructions confine the entangling action to a small number of targeted two-qubit blocks. As a result, when deploying the circuits on restricted coupling graphs, the need for additional SWAP routing is often substantially reduced in practice~\cite{Li2019,QiskitSabreSwap}.

Finally, our circuits can reduce resources relative to universal three-qubit constructions when additional structure is present~\cite{Plesch2011}. A quantitative comparison with the fixed-depth template of \cite{Giraud2009} is summarized in Table~\ref{tab:resource_comparison}. Across the $\mathrm{R}_k$ families considered here, our class-specific circuits consistently achieve smaller gate counts and shallower overall depth than the fixed template, with particularly pronounced savings for the $\mathrm{R}_1$ and $\mathrm{R}_2$ classes. Such reductions are especially valuable in near-term, NISQ-era implementations, where CNOT-gate count and overall circuit depth are primary contributors to infidelity~\cite{Preskill2018}.

A natural direction for future work is to understand to what extent this Schmidt-decomposition approach can be made scalable beyond three qubits, in light of general depth and resource scaling considerations for $n$-qubit state preparation~\cite{Zhang2022,Preskill2018}. Beyond improving performance on NISQ hardware, it is also of interest to explore how the same ideas might be adapted to FTQC settings, where error-correction overheads can amplify the cost of circuit depth and entangling operations~\cite{Preskill2018}. One promising route is to retain the same organizing principle—choosing a fixed bipartition and extracting bipartite structural data—while developing a hierarchy of compilation modules that systematically lifts an $n$-qubit preparation routine to an $(n+1)$-qubit routine. This raises several concrete questions: how to control the growth of two-qubit-gate count under iterative bipartition-based synthesis, how to preserve hardware-aware placement as the system size increases, and which structured state families admit class-specific simplifications analogous to the $\mathrm{R}_k$ classes. Addressing these issues would clarify the practical scalability of the framework and may lead to a general constructive methodology for deterministic state preparation with explicit parameters and deployable connectivity.

\begin{acknowledgments}
We are grateful to Soojoon Lee for his valuable assistance and insightful discussions.
This research was supported by Basic Science Research Program through the National Research Foundation of Korea (NRF) funded by the Ministry of Education (Grant No. NRF-2020R1I1A1A01058364, Grant No. RS-2023-00243988, Grant No. RS-2025-25415913, and Grant No. RS-2024-00432600).
\end{acknowledgments}


\nocite{*}

\bibliography{SP}

\end{document}